\documentclass[12pt]{article}
\usepackage{epsf}
\usepackage{a4wide,graphicx}
\usepackage{cite}
\begin{document}

\title{$VP\gamma$ radiative decay of resonances dynamically generated from the vector meson-vector meson interaction}

\author{J. Yamagata-Sekihara$^1$ and E. Oset$^1$ \\
{\small{\it $^1$Departamento de F\'{\i}sica Te\'orica and IFIC,
Centro Mixto Universidad de Valencia-CSIC,}}\\
{\small{\it Institutos de
Investigaci\'on de Paterna, Aptdo. 22085, 46071 Valencia, Spain}}\\
}

\date{\today}

\maketitle

 \begin{abstract}
We evaluate the radiative decay into a vector a pseudoscalar and a 
photon of several resonances dynamically generated from the vector 
vector interaction. The process proceeds via the decay of one of the 
vector components into a pseudoscalar and a photon, which have an 
invariant mass distribution very different from phase space as a 
consequence of the two vector structure of the resonances. Experimental 
work along these lines should provide useful information on the nature 
of these resonances.
\end{abstract}

\section{Introduction}

   The success of the chiral unitary approach generating resonances from the interaction of pseudoscalar mesons or pseudoscalar mesons with baryons, providing the properties of these resonances and their production cross sections in different reactions \cite{review,puri} has stimulated work replacing pseudoscalar mesons by vector mesons. 
A natural extension of the chiral Lagrangians to incorporate vector mesons and
their interaction is provided by the hidden local gauge formalism for vector
interactions with pseudoscalar
mesons, vectors and photons \cite{hidden1,hidden2,hidden3,hidden4}.
 Once again the skilful combination of the interaction provided
by these Lagrangians with unitary techniques in coupled channels allows one
to obtain a realistic approach to study the vector-vector interaction, and
  work in this direction has been already done in 
\cite{raquel,geng} studying the vector-vector interaction up to about 2000 MeV. The nonperturbative unitary
techniques are essential there and  many resonances are generated within the scheme. 
 In practice one solves a set of coupled channels Bethe
Salpeter equations using as Kernel the interaction provided by the hidden gauge
Lagrangians, regularizing the loops with a natural scale \cite{ollerulf}. Several mesonic resonances are found as poles in the scattering matrices, indicating that one has some kind of molecular states.  Actually,
there are strong experimental arguments to suggest that the  $f_0(1370)$ is a $\rho \rho$ molecule \cite{klempt,crede}.
The
results of \cite{raquel} show that the $f_0(1370)$ and $f_2(1270)$ mesons are
dynamically generated from the $\rho \rho$ interaction. 

The work of
\cite{geng} extends that of \cite{raquel} to the interaction of all members of the vector
nonet resulting in the dynamical generation of eleven resonances, 
some of which can be associated
to known resonances ($f_0(1370)$, $f_2(1270)$, $f'_2(1525)$, $f_0(1710)$ and 
 $K^*_2(1430)$), while others are predictions.

 The nature of these resonances as molecular states of a pair of vector mesons,  allows one to evaluate many observables like the radiative decay of the 
$f_0(1370)$ and $f_2(1270)$ mesons into $\gamma \gamma$ \cite{yamagata}, were
agreement with the experimental data is found. Similarly, the $J/\psi$ decay into $\phi (\omega)$ and one of those resonances \cite{chinacola}, and the $J/\psi$ radiative decay into
$\gamma$ and one of the those resonances \cite{chinavalgerman}, were also found consistent
with experiment.

The idea of vector-vector molecules has also found support in alternative studies.  In \cite{gutsche} the Y(3940) and Y(4140) are assumed to be bound states of $D^*$ and $\bar{D}^*$ and  $D^*_s$ and $\bar{D}^*_s$ respectively and the Weinberg compositness condition \cite{weinberg1,weinberg2,Hanhart:2007yq,Baru:2003qq} is invoked to get the coupling of the Y resonances to these components. Another approach,
based on chiral symmetry and heavy quark symmetry has
been used in \cite{shilinzhu}, where also bound states of the
$D^* \bar{D^*}$ systems are found in some cases.

   The particular structure of these states induced the idea \cite{liuke} that the study of their decay into $D^*\bar{D}\gamma$, or $D^*_s\bar{D_s} \gamma$ should provide o good test for the claimed structure of these resonances. 
   The idea was caught up in \cite{weihong} and applied to several X,Y,Z resonances  of hidden charm nature which are dynamically generated within the hidden gauge approach from the interaction of vector mesons \cite{xyz}.
   
    In the present paper we shall further pursue this idea for the states generated in \cite{raquel,geng} and study the decay of the 
$f_0(1370)$, $f_2(1270)$, $f'_2(1525)$, $f_0(1710)$ and the $K^*_2(1430)$ resonances into several channels involving one vector, one pseudoscalar and a photon. While we find similar results for the shapes of the invariant mass distributions of the pseudoscalar-photon pair as in \cite{weihong}, we find that the decay rates obtained in this case are far larger than those obtained in the charm sector. The experimental investigation of these decay rates would provide further information concerning the claimed nature of those resonances as vector-vector molecules, and provide further support for the similar nature claimed for some of the X,Y,Z resonances \cite{gutsche,shilinzhu,xyz}.
  
\section{Formalism}
In \cite{geng} the $f_0(1370),~f_0(1710),~f_2(1270),~f_2'(1520)$, and $K^*_2(1430)$ were dynamically generated by the vector-vector interaction.
The states were identified by observing poles in the vector-vector scattering matrix with certain quantum numbers.  The real part of
the pole position provides the mass of the resonance and the imaginary part one
half its width. In addition the residues at the poles provide the product of the
coupling of the resonance to the initial and final channels, from where, by
looking at the scattering amplitudes in different channels, we can obtain the
coupling of the resonance to all channels up to an irrelevant global sign for
just one coupling.
In Table~\ref{tab:1} the couplings to the most relevant channel are shown.

\begin{table}[htbp]
\begin{center}
\caption{\label{tab:1}Coupling constans of main decay channel of the resonances of ~\cite{geng}. All quantities are in units of MeV.}
\begin{tabular}{l|c|ccc}
\hline
\hline
Resonance&spin&\multicolumn{3}{c}{coupling constant $g$}\\\hline
$f_0(1370)$&J=0&$ \rho \rho $&&\\
$$&&(7920,-$i$1071)&&\\ \hline
$f_0(1710)$&J=0&$K^*{\bar K^*}$& $\phi \omega $&$\omega \omega $ \\
$$&&(7124,$i$96)&(3010,-$i$210)&(-1763,$i$108)\\ \hline
$f_2(1270)$&J=2&$\rho \rho$&&\\
$$&&(10889,-$i$99)&&\\ \hline
$f_2'(1520)$&J=2&$K^*{\bar K^*}$&$\phi \omega$&$\omega \omega$\\
$$&&(10121,$i$101)&(5016,-$i$17)&(-2709,$i$8)\\ \hline
$K^*_2(1430)$&J=2&$\rho K^*$&$K^*\omega$&\\
$$&&(10901,-$i$71)&(2267,-$i$13)&\\ \hline
\hline
\end{tabular}
\end{center}
\end{table} 

In \cite{geng}  these couplings are given in isospin basis.
   However, we need them now in charge basis, which are readily obtained for the
   isospin combinations
\begin{eqnarray}
|\rho\rho,I=0\rangle &=&-\frac{1}{\sqrt{3}}(|\rho^{+}\rho^{-}\rangle +|\rho^-\rho^+\rangle+
|\rho^0\rho^0\rangle),\nonumber \\
|K^*K^*,I=0\rangle &=&-\frac{1}{\sqrt{2}}(|K^{*+}K^{*-}\rangle -
|K^{*-}K^{*+}\rangle)~~.
\end{eqnarray}
In addition, the couplings of \cite{geng} are calculated with the unitary normalization (extra $1/\sqrt{2}$ factor to account for identical particles in the sum over intermediate states).
Thus, the couplings of $\rho\rho$, $K^*{\bar K^*}$ and $\omega\omega$ components must be multiplied by $(\sqrt{2/3})$, $1$ and $\sqrt{2}$ to get the appropriate coupling for the charged or neutral states
   (a sign is irrelevant for the width).

\begin{figure}[htpd]
\begin{center}
\includegraphics[width=5.5cm,height=4.5cm]{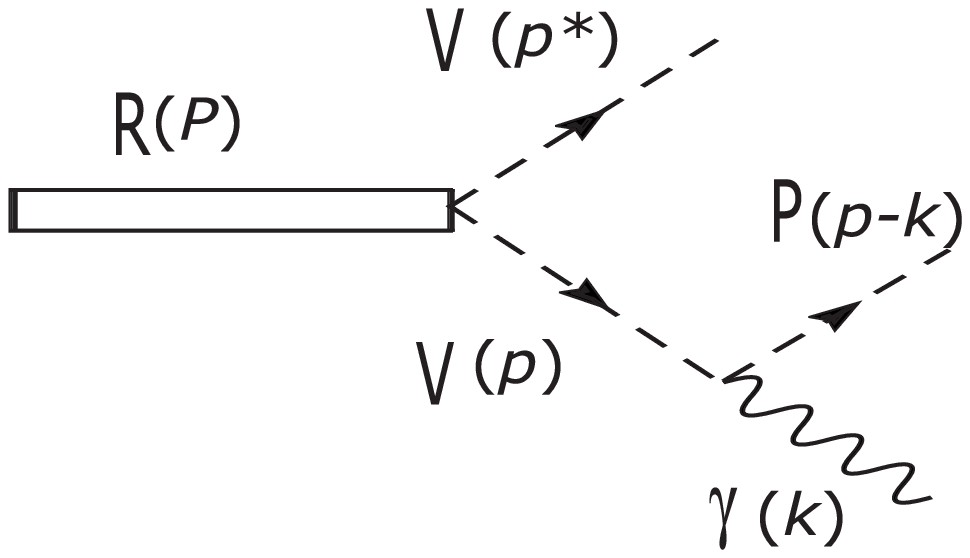}
\caption{\label{fig:diagram}}
\end{center}
\end{figure}

In the present work we address the problem of the decay  mode of the resonance when one vector meson further decays into a pseudoscalar meson and a photon.
The corresponding Feynman diagram is shown in Fig.~\ref{fig:diagram}.

 The spin projection operators on $J=0,2$, evaluated assuming the 
 three momenta of $V$ to be small with respect to the mass of the vector
 mesons, which is indeed the case here, are given in terms of the polarization
 vectors by
\begin{eqnarray}
P'^{(0)}&=&\frac{1}{\sqrt{3}}\epsilon_i^{(1)}\epsilon_i^{(2)},\nonumber \\
P'^{(2)}&=&\left\{\frac{1}{2}\left(\epsilon_i^{(1)}\epsilon_j^{(2)}+\epsilon_j^{(1)}\epsilon_i^{(2)}\right)-\frac{1}{3}\epsilon_l^{(1)}\epsilon_l^{(2)}\delta_{ij}\right\},
\label{eq:projectors}
\end{eqnarray}
 where $i,j$ are spatial indices. On the other hand the anomalous vertex for the $V$ decay
 into $P \gamma$ is given by
\begin{equation}
-it_{V \to P \gamma}=-ig_{V\gamma P
}\epsilon_{\mu\nu\alpha\beta}p^{\mu}\epsilon^{\nu}(V)k^{\alpha}\epsilon^{\beta}(\gamma),
\end{equation}
which gives rise to a width
\begin{equation}
\Gamma_{V \to P \gamma}=\frac{1}{48\pi}g^2_{V\gamma P
}\frac{k}{M^2_{V}}(M^2_{V}-m^2_{P})^2.
\label{eq:radwidth}
\end{equation}
Using Eq.~(\ref{eq:radwidth}) and values of the PDG for the vector radiative widths, we obtain the coupling of $g_{VP\gamma}$
 for the $V\to P \gamma$ decay, which are given by
 \begin{eqnarray}
 g_{\rho^\pm \to \pi^\mp \gamma}&=& 2.19\times10^{-4}~{\rm MeV}^{-1}\nonumber\\
 g_{\rho^0 \to \pi0 \gamma}&=& 2.52\times10^{-4}~{\rm MeV}^{-1}\nonumber\\
 g_{K^{*\pm} \to K^\pm \gamma}&=& 2.53\times10^{-4}~{\rm MeV}^{-1}\nonumber\\
 g_{K^{*0} \to K^0 \gamma}&=& 2.19\times10^{-4}~{\rm MeV}^{-1}\nonumber\\
 g_{\omega \to \pi^0 \gamma}&=& 6.96\times10^{-4}~{\rm MeV}^{-1}\nonumber~~.
\end{eqnarray}

Let us begin with the $f_0(1370)$ case.
   With the previous information we can already write the amplitude for the
   decay of the $f_0(1370)$ into $\rho^+\pi^- \gamma$, which is given by
\begin{eqnarray}
-it&=&-i\frac{\sqrt{2}}{\sqrt{3}}\tilde{g}\frac{1}{\sqrt{3}}\epsilon_i^{(1)}\epsilon_i^{(2)}\frac{i}{p^2-M^2_{\rho}+iM_{\rho}\Gamma_{\rho}}\nonumber \\
&&\times (-i)g_{\rho\pi\gamma}\epsilon_{\mu\nu\alpha\beta}p^{\mu}\epsilon^{\nu
(2)}k^{\alpha}\epsilon^{\beta}(\gamma),
\end{eqnarray}
   where the indices (1), (2) indicate the $\rho^+$ and the $\rho^-$ respectively.
   The sum over the intermediate $\rho^-$ polarizations can be readily done as
 \begin{equation}
 \sum\limits_{\lambda}\epsilon_i^{(2)}\epsilon^{\nu
 (2)}=-g_i^\nu=-\delta_{i\nu},
 \end{equation}
   where we have neglected the three momenta of the intermediate $\rho^-$ which
   is in average very small compared with the $\rho^-$ mass, particularly at
   large invariant masses of the $\pi^- \gamma$ system which concentrates most
   of the strength, as we shall see. The sum  of $|t|^2$ over the final
   polarizations of the vector and the photon is
   readily done and, neglecting again terms of order $\vec{p}\,^2/M_{\rho}^2$,
    we get the  result
\begin{eqnarray}
\sum|t|^2&=&\frac{2}{3}\frac{1}{3}\tilde{g}^2g^2_{\rho\pi\gamma}\left|\frac{1}{p^2-M^2_{\rho}+iM_{\rho}\Gamma_{\rho}}\right|^2
2(p\cdot k)^2\nonumber \\
 &=&\frac{2}{9}\frac{1}{2}\tilde{g}^2g^2_{\rho\pi\gamma}\left|\frac{p^2-m^2_\pi}{p^2-M^2_{\rho}+iM_{\rho}\Gamma_{\rho}}\right|^2.
\label{eq:tdos}
 \end{eqnarray}

   The differential mass distribution with respect to the  invariant mass of the
   $\rho^{-} \gamma$ system, which is equal to $p^2$, is finally given by
\begin{equation}
\frac{d
\Gamma_R}{dM_{\rm inv}}=\frac{1}{4M_R^2}\frac{1}{(2\pi)^3}p_{\rho}
\tilde{p}_\pi\sum|t|^2,
\label{eq:dgamma}
\end{equation}
   where $p_{\rho}$ is the momentum of the $\rho^{+}$ in the rest frame of the
   resonance $R$ and $\tilde{p}_\pi$ is the momentum of the $\pi^-$ in the rest
   frame of the final  $\pi^{-} \gamma$ system given by
\begin{eqnarray}
p_{\rho}&=&\frac{\lambda^{1/2}(M_R^2,M^2_{\rho},M^2_{\rm inv})}{2M_R},\nonumber \\
\tilde{p}_\pi&=&\frac{M^2_{\rm inv}-m^2_\pi}{2M_{\rm inv}}.
\end{eqnarray}

One further step must be taken since the resonance and vector meson have decay widths. To take this into account, we consider the mass distribution of these two states and convolute the expression of the width with the mass distribution of the two particles.
The differential mass distribution of the radiative decay of the resonance is then given as,
\begin{equation}
(\frac{d\Gamma_R}{dM_{\rm inv}})'=\frac{F}{G}~~,
\end{equation}
where $F$ and $G$ are given by
\begin{eqnarray}
F&=&\int^{m_\rho+\Gamma_\rho/2}_{m_\rho-\Gamma_\rho/2}d\tilde{m_\rho}(-\frac{1}{\pi}){\rm Im}\frac{1}{{\tilde m_\rho}^2-m_\rho^2+i\Gamma_{\rho}m_\rho}\nonumber\\
&\times&\int^{M_R+\Gamma_R/2}_{M_R-\Gamma_R/2}(-\frac{1}{\pi}){\rm Im}\frac{1}{{\tilde M_R}^2-M_R^2+i\Gamma_RM_R}\times\frac{d\Gamma_R}{dM_{\rm inv}}({\tilde m_\rho},{\tilde M_R},M_{\rm inv})\\
G&=&\int^{m_\rho+\Gamma_\rho/2}_{m_\rho-\Gamma_\rho/2}d\tilde{m_\rho}(-\frac{1}{\pi}){\rm Im}\frac{1}{{\tilde m_\rho}^2-m_\rho^2+i\Gamma_{\rho}m_\rho}\nonumber\\
&\times&\int^{M_R+\Gamma_R/2}_{M_R-\Gamma_R/2}(-\frac{1}{\pi}){\rm Im}\frac{1}{{\tilde M_R}^2-M_R^2+i\Gamma_RM_R}
\end{eqnarray}

   In the case of the tensor state we must do extra work since the
projector operators are different. In this case we must keep the indices $i, j$
in $t$ and multiply with $t^*$ with the same indices $i, j$. This sums over all
possible final polarizations but also the initial $R$ polarizations, so in order
to take the sum and average over final and initial polarizations, respectively,
one must divide the results of the $i,j$ sum of $tt^*$ by $(2J+1)$, where $J$ is
the spin of the resonance R.  The explicit evaluation for the case of the tensor
states, $J=2$, of $\rho\rho$ proceeds as follows: The $t$ matrix is now
written as
\begin{eqnarray}
t&=&\frac{1}{\sqrt{2}}\tilde{g}g_{\rho\pi\gamma} \left\{
\frac{1}{2}\left(\epsilon_i^{(1)}\epsilon_j^{(2)}+\epsilon_j^{(1)}\epsilon_i^{(2)}
\right)
-\frac{1}{3}\epsilon_l^{(1)}\epsilon_l^{(2)}\delta_{ij}\right\}\nonumber \\
&&\times\frac{1}{p^2-M^2_{\rho}+iM_{\rho}\Gamma_{\rho}}\epsilon_{\mu\nu\alpha\beta}p^{\mu}\epsilon^{\nu
(2)} k^\alpha\epsilon^\beta(\gamma).
\end{eqnarray}

 As mentioned above, we must multiply $t_{i,j}$ by $t^*_{i,j}$, recalling that
 the indices $i,j$ are spatial indices and divide by $(2J+1)$ (5 in this case)
 in order to obtain the modulus squared of the transition matrix, summed and
 averaged over the final and initial polarizations.  Neglecting again terms
 that go like $\vec{p}\,^2/m_\rho$ we obtain the same expression as in 
 Eq. (\ref{eq:tdos}).

\section{Results}

We show here the results for different cases:

\subsection{$f_0(1370)$ and $f_0(1710)$}
The decay modes considered are
\begin{eqnarray}
f_0(1370)&\to&\rho\pi\gamma\nonumber\\
f_0(1710)&\to&K^*{\bar K}\gamma\nonumber\\
f_0(1710)&\to&\phi\pi\gamma\\
f_0(1710)&\to&\omega\pi\gamma\nonumber~~.
\label{eq:f0}
\end{eqnarray}

\begin{figure}[htpd]
\begin{center}
\includegraphics[width=8.cm,height=6.0cm]{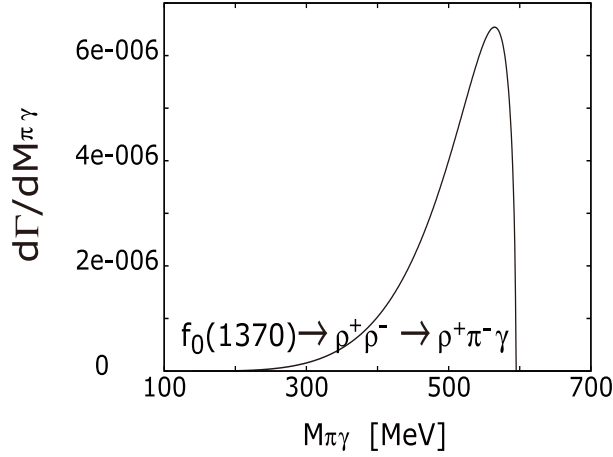}
\caption{\label{fig:noconv}Differential mass distribution for $f_0(1370)\to\rho^+\pi^-\gamma$ without convolution.}
\end{center}
\end{figure}

In Fig.~\ref{fig:noconv} we show the calculated differential mass distribution of $f_0(1370)\to\rho^+\pi^-\gamma$.
This calculation has done without the decay widths of $f_0(1370)$ and $\rho^+$ meson, namely Eq.~(\ref{eq:dgamma}).
Since the $f_0(1370)$ resonance is below the $\rho^+\rho^-$ threshold, the peak position appears below the $\rho^-$ mass energy.

\begin{figure}[htpd]
\begin{center}
\includegraphics[width=13.cm,height=9.0cm]{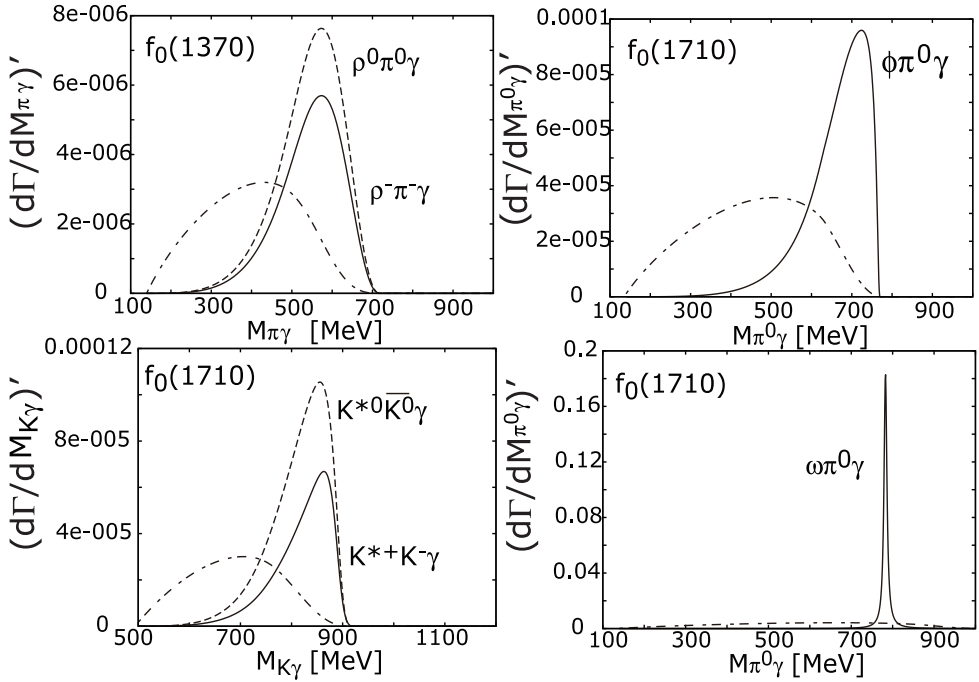}
\caption{\label{fig:I0}Differential mass distribution for the case of $f_0(1370)$ and $f_0(1710)$.
The phase space calculation corresponds to the decay channel of the solid line with the same normalization.}
\end{center}
\end{figure}

In Fig.~\ref{fig:I0}, we consider the decay widths of the resonance and vector meson.
For the $f_0(1370)$ resonance the width is poorly determined  experimentally. We take a slice of the resonance around its peak and use   $\Gamma_{f_0(1370)}=100~{\rm MeV/c}$. With this window we prevent that the resonance goes into two physical $\rho$ mesons, one of which would decay into $\pi \gamma$.
For the  $f_0(1710)$  the decay into two vector mesons is forbidden except in the case of $\omega \omega$ final state. This is the reason why the decay rate into $\omega \pi^0 \gamma$ is exceptionally large, since it corresponds to the decay of the resonance into $\omega \omega$ followed by the decay of either of the two $\omega$ into $\pi^0 \gamma$. The phase space is not restricted, unlike  in the other cases where the intermediate vector meson that decays into a pseudoscalar meson and a photon is necessarily off shell. In the figure we compare the mass distribution with
 the phase space distribution of each decay mode (dashed dotted line), obtained omitting the $p^2$ dependent terms in eq. (\ref{eq:tdos}).  As we can see, the shapes obtained with the dynamical picture of the resonances is very different from what one gets using simple phase space.\\

\subsection{$f_2(1270)$ and $f_2'(1520)$ case}
\begin{figure}[htpd]
\begin{center}
\includegraphics[width=13.cm,height=9.0cm]{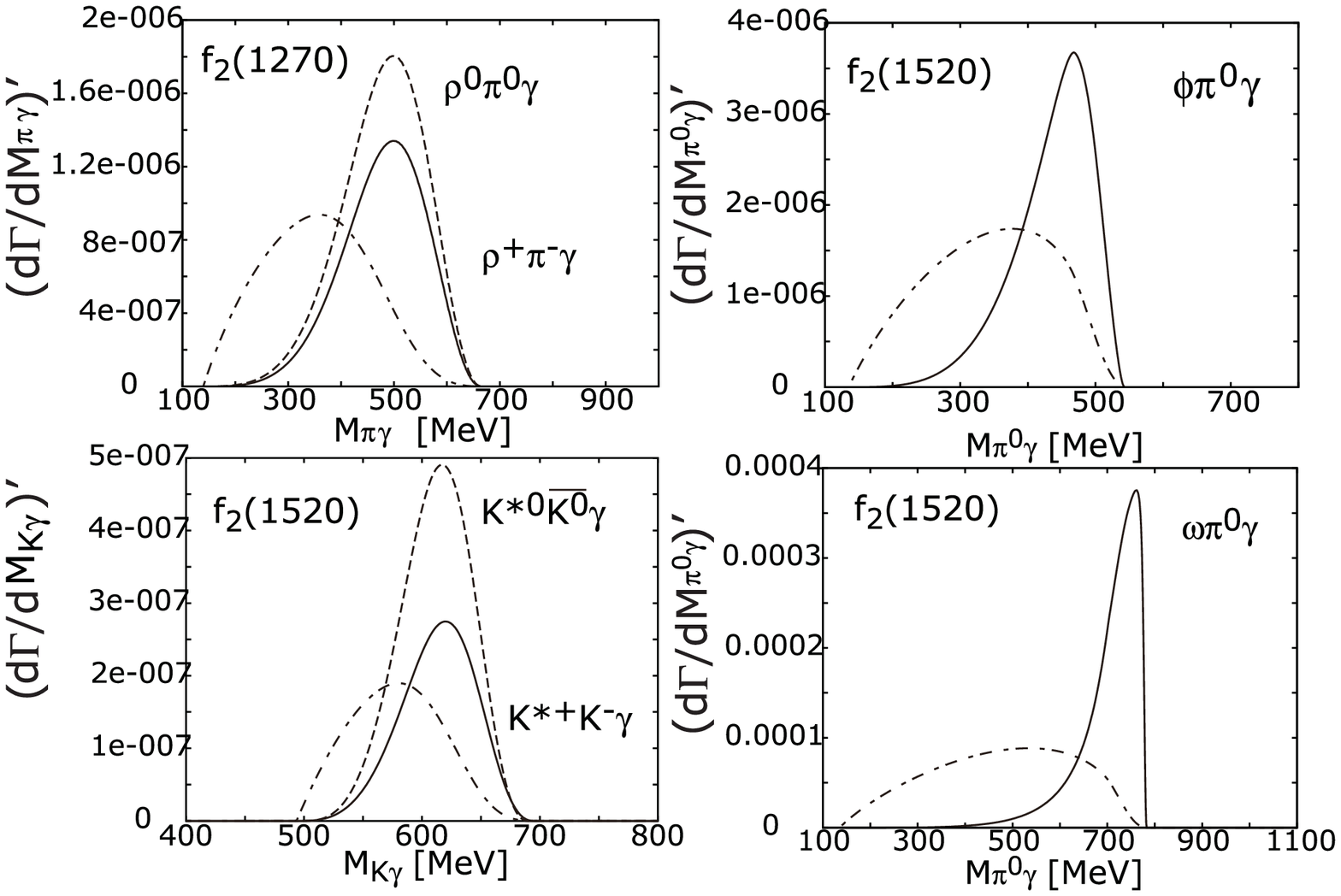}
\caption{\label{fig:I2}Differential mass distribution for the case of $f_2(1270)$ and $f_2(1520)$.
The phase space calculation corresponds to the decay channel of the solid line with the same normalization.}
\end{center}
\end{figure}

The decay modes considered are:
\begin{eqnarray}
f_2(1270)&\to&\rho\pi\gamma\nonumber\\
f_2'(1520)&\to&K^*{\bar K}\gamma\nonumber\\
f_2'(1520)&\to&\phi\pi\gamma\\
f_2'(1520)&\to&\omega\pi\gamma\nonumber~~.
\label{eq:f2}
\end{eqnarray}
The $f_2(1270)$ and $f_2'(1520)$ decay invariant mass distributions are calculated as before and the results are shown in Fig.~\ref{fig:I2}. Once again we see the striking difference between the results obtained from the dynamical picture of the resonances and phase space (dashed dotted lines).

\subsection{$K_{2}^*(1430)$ case}
\begin{figure}[htpd]
\begin{center}
\includegraphics[width=13.cm,height=9.0cm]{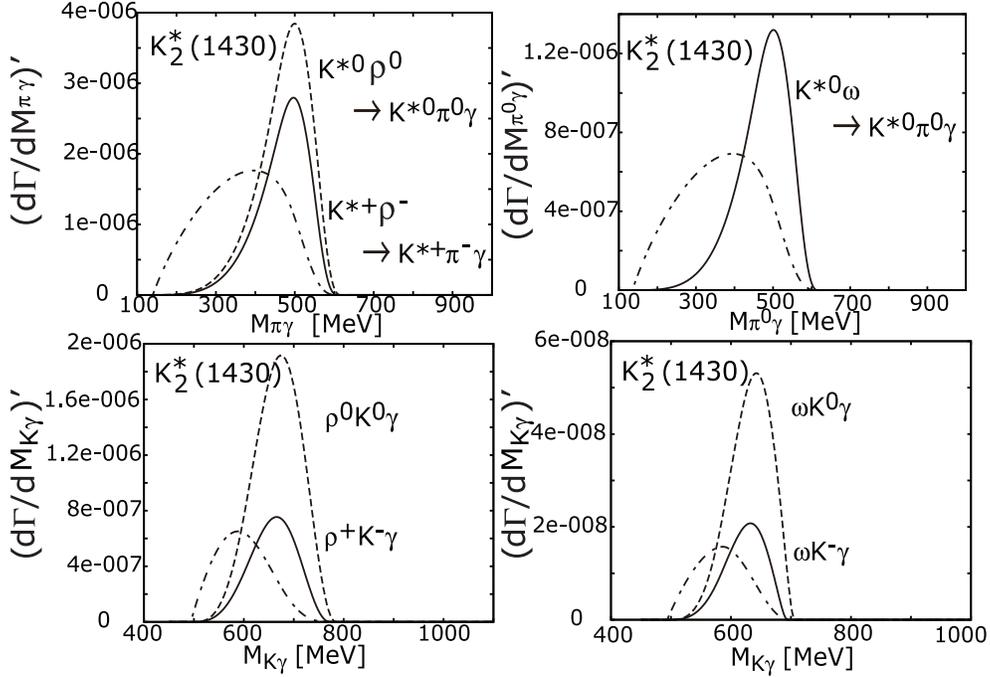}
\caption{\label{fig:I2K}Differential mass distribution for the case of $K^*_2(1430)$.
The phase space calculation corresponds to the decay channel of the solid line with the same normalization.}
\end{center}
\end{figure}

The decay modes considered are:
\begin{eqnarray}
K^*_2(1430)&\to&K^{*}\rho\to K^{*}\pi\gamma\nonumber\\
K^*_2(1430)&\to&\rho K\gamma\nonumber\\
K^*_2(1430)&\to&K^{*}\omega\to K^{*}\pi\gamma\\
K^*_2(1430)&\to&\omega K\gamma\nonumber~~.
\label{eq:f2_K}
\end{eqnarray}
We show the results of the $K^*_2(1430)$ decay in Fig.~\ref{fig:I2K}, where once again we see the striking differences with the results obtained with those with just phase space.

In Table~\ref{tab:gamma}, we show the results of the width of vector-vector meson $\Gamma_{R\to VV}$ and the radiative decay width $\Gamma_{R\to VP\gamma}$
The radiative decay width $\Gamma_{R\to VP\gamma}$ is obtained  integrating the differential mass distribution.

  The large radiative decay width of the $f_0(1710) \to \omega \pi^0 \gamma$ comes, as indicated before, because the decay into $\omega \omega $ is now allowed. This case serves us to make a test of the calculation. The radiative decay width should be twice the product of the $f_0(1710) \to
  \omega  \omega$ width times the branching ratio of the $\omega \to \pi^0 \gamma$, which is  experimentally  8.28 \%.
\begin{equation}
\Gamma_{f_0(1710)\to \omega\pi^0\gamma}=\Gamma_{f_0\to \omega\omega}\times 2\times B_{\omega\to\pi^0\gamma}
\label{eq:a}
\end{equation}
with
\begin{equation}
\Gamma_{f_0\to\omega\omega}=\frac{1}{8\pi M^2_{f_0}}g^2k
\label{eq:b}
\end{equation}
where $k$ is the $\omega$ momentum in the $f_0((1710)\to \omega\omega$ decay, and $g$ is the $f_0(1710)$ coupling to $\omega\omega$ from \cite{geng}, shown in Table~\ref{tab:1}.
This coupling incorporates the unitary normalization of the $\omega\omega$ (extra factor $1/\sqrt{2}$) which makes unnecessary to divide by a factor of two the $f_0(1710)\to \omega\omega$ width in Eq.~(\ref{eq:b}).
The factor 2 in Eq.~(\ref{eq:a}) accounts for the $\omega\to \pi^0\gamma$ decay of each of the two omegas.
The result of Table~\ref{tab:gamma} for $f_0(1710)\to\omega\pi^0\gamma$ fulfills the relationship of Eq.~(\ref{eq:a}).

\begin{table}[htbp]
\begin{center}
\caption{\label{tab:gamma} Radiative decay width of vector-vector meson $\Gamma_{R\to VP\gamma}$.}
\begin{tabular}{l|c}
\hline
\hline
Decay mode&$\Gamma_{R\to VP\gamma}$~[keV]\\\hline
$f_0(1370)\to \rho^\pm\rho^\mp\to\rho^\pm\pi^\mp\gamma$&1.06\\
$f_0(1370)\to \rho^0\rho^0\to \rho^0\pi^0\gamma$&1.42\\
$f_0(1710)\to K^{*\pm}K^{*\mp}\to K^{*\pm}K^{\mp}\gamma$&7.30\\
$f_0(1710)\to K^{*0}{\bar K^{*0}}\to K^{*0}{\bar K^0}\gamma$&12.73\\
$f_0(1710)\to \phi\omega\to \phi\pi^0\gamma$&14.45\\
$f_0(1710)\to \omega\omega\to \omega\pi^0\gamma$&2.40$\times 10^3$\\
$f_2(1270)\to \rho^\pm\rho^\mp\to\rho^\pm\pi^\mp\gamma$&2.63$\times 10^{-1}$\\
$f_2(1270)\to \rho^0\rho^0\to\rho^0\pi^0\gamma$&3.55$\times 10^{-1}$\\
$f_2'(1520)\to K^{*\pm}K^{*\mp}\to K^{*\pm}K^{\mp}\gamma$&$2.18\times 10^{-2}$\\
$f_2'(1520)\to K^{*0}{\bar K^{*0}}\to K^{*0}{\bar K^0}\gamma$&$3.77\times 10^{-2}$\\
$f_2'(1520)\to \phi\omega\to \phi\pi^0\gamma$&$4.60\times 10^{-1}$\\
$f_2'(1520)\to \omega\omega\to \omega\pi^0\gamma$&$38.16$\\
$K^{*}_2(1430)\to K^*\rho^\pm \to K^*\pi^\pm\gamma$&$4.07\times10^{-1}$\\
$K^{*}_2(1430)\to K^*\rho^0 \to K^*\pi^0\gamma$&$5.81\times10^{-1}$\\
$K^{*}_2(1430)\to \rho^+K^{*-} \to \rho^+K^-\gamma$&$8.99\times10^{-2}$\\
$K^{*}_2(1430)\to \rho^0{\bar K^{*0}} \to \rho^0{\bar K^0}\gamma$&$2.36\times10^{-1}$\\
$K^{*}_2(1430)\to K^*\omega \to K^*\pi^0\gamma$&$1.99\times10^{-1}$\\
$K^{*}_2(1430)\to \omega K^{*-} \to \omega K^-\gamma$&$1.78\times10^{-3}$\\
$K^{*}_2(1430)\to \omega{\bar K^{*0}} \to \omega{\bar K^0}\gamma$&$4.79\times10^{-3}$\\
\hline
\end{tabular}
\end{center}
\end{table}

\section{Summary}
We have studied the radiative decay width of $f_0(1370)$, $f_0(1710)$, $f_2(1250)$, $f_2'(1520)$ and $K^*(1430)$ resonances, which are dynamically generated by the vector-vector interaction, into a vector, a pseudoscalar and a photon.  Except in one case, the 
$f_0(1710) \to \omega \pi^0 \gamma$, where all the strength of the $\pi^0 \gamma$ invariant mass accumulates at the $\omega$ mass value, in all the other cases we find wider distributions, quite different from what one expects in terms of phase space. The memory of the resonance as been a bound state of two vector mesons is responsible for this shape, and the strength of the pseudoscalar plus photon invariant mass acumulates as close to the mass of the vector meson as possible, within the phase space availability. The case of the $f_0(1710) \to \omega \pi^0 \gamma$ is special because what one sees is the decay into $\omega \omega$. Since the branching ratio of the $\omega$ to $\pi^0 \gamma$ is relatively big (it is actually used to detect $\omega$ in the TAPS detector
\cite{mariana}), this decay mode would be a direct measurement of the 
$f_0(1710) \to \omega \omega$ decay, allowing to test predictions of the theoretical framework of the vector-vector coupled channels approach.

   The rates obtained for the radiative decays are relatively large in some cases, of the order of the keV or tens of keV. These magnitudes are easily measurable and the results obtained here should stimulate experimental work in this direction, which should teach us much regarding the nature of the resonances studied.

\section*{Acknowledgments}
This work is partly supported by DGICYT contract number FIS2006-03438 and the Generalitat Valenciana in the Program Prometeo.
We acknowledge the support of the European Community-Research Infrastructure Integrating Activity Study of Strongly Interacting Matter (acronym HadronPhysics2, Grant Agreement n. 227431) under the Seventh Framework Programme of EU.
J. Y. is the Yukawa Fellow and this work is partially supported by Yukawa Memorial Foundation.


\begin{thebibliography}{99}

\bibitem{review}
  J.~A.~Oller, E.~Oset and A.~Ramos,
  Prog.\ Part.\ Nucl.\ Phys.\  {\bf 45}, 157 (2000)
  
\bibitem{puri}
  E.~Oset, D.~Cabrera, V.~K.~Magas, L.~Roca, S.~Sarkar, M.~J.~Vicente Vacas and A.~Ramos,
  Pramana {\bf 66}, 731 (2006)
  [arXiv:nucl-th/0504033].

 

\bibitem{hidden1}
  M.~Bando, T.~Kugo, S.~Uehara, K.~Yamawaki and T.~Yanagida,
  Phys.\ Rev.\ Lett.\  {\bf 54}, 1215 (1985).


\bibitem{hidden2}
  M.~Bando, T.~Kugo and K.~Yamawaki,
  Phys.\ Rept.\  {\bf 164}, 217 (1988).

\bibitem{hidden3}
  M.~Harada and K.~Yamawaki,
  Phys.\ Rept.\  {\bf 381}, 1 (2003)
  
\bibitem{hidden4}
  U.~G.~Meissner,
  Phys.\ Rept.\  {\bf 161}, 213 (1988).


     
\bibitem{raquel}
  R.~Molina, D.~Nicmorus and E.~Oset,
  Phys.\ Rev.\  D {\bf 78}, 114018 (2008)
  
\bibitem{geng}
  L.~S.~Geng and E.~Oset,
  Phys.\ Rev.\  D {\bf 79}, 074009 (2009)
  

\bibitem{ollerulf}
  J.~A.~Oller and U.~G.~Meissner,
  Phys.\ Lett.\  B {\bf 500}, 263 (2001)
    
     
\bibitem{klempt}
  E.~Klempt and A.~Zaitsev,
  Phys.\ Rept.\  {\bf 454}, 1 (2007)


\bibitem{crede}
  V.~Crede and C.~A.~Meyer,
  Prog.\ Part.\ Nucl.\ Phys.\  {\bf 63}, 74 (2009)








\bibitem{yamagata}
  H.~Nagahiro, J.~Yamagata-Sekihara, E.~Oset, S.~Hirenzaki and R. Molina,
  Phys.\ Rev.\  D {\bf 79}, 114023 (2009)


\bibitem{chinacola}
  A.~Martinez Torres, L.~S.~Geng, L.~R.~Dai, B.~X.~Sun, E.~Oset and B.~S.~Zou,
  Phys.\ Lett.\  B {\bf 680}, 310 (2009)

\bibitem{chinavalgerman}  
  L.~S.~Geng, F.~K.~Guo, C.~Hanhart, R.~Molina, E.~Oset and B.~S.~Zou,
  arXiv:0910.5192 [hep-ph].
  
\bibitem{gutsche}
  T.~Branz, T.~Gutsche and V.~E.~Lyubovitskij,
  Phys.\ Rev.\  D {\bf 80}, 054019 (2009)
  
\bibitem{weinberg1}
  S.~Weinberg,
  Phys.\ Rev.\  {\bf 130}, 776 (1963).
  
\bibitem{weinberg2}
  S.~Weinberg,
  Phys.\ Rev.\  {\bf 137}, B672 (1965).



\bibitem{Hanhart:2007yq}
  C.~Hanhart, Yu.~S.~Kalashnikova, A.~E.~Kudryavtsev and A.~V.~Nefediev,
  Phys.\ Rev.\  D {\bf 76}, 034007 (2007)
  [arXiv:0704.0605 [hep-ph]].

\bibitem{Baru:2003qq}
  V.~Baru, J.~Haidenbauer, C.~Hanhart, Yu.~Kalashnikova and A.~E.~Kudryavtsev,
  Phys.\ Lett.\  B {\bf 586}, 53 (2004)
  [arXiv:hep-ph/0308129].
  
\bibitem{shilinzhu}
  X.~Liu, Z.~G.~Luo, Y.~R.~Liu and S.~L.~Zhu,
  Eur.\ Phys.\ J.\  C {\bf 61}, 411 (2009)
  
\bibitem{liuke}
  X.~Liu and H.~W.~Ke,
  Phys.\ Rev.\  D {\bf 80}, 034009 (2009)
  [arXiv:0907.1349 [hep-ph]].
  
\bibitem{weihong}
  W.~H.~Liang, R.~Molina and E.~Oset,
  arXiv:0912.4359 [hep-ph].


\bibitem{xyz}
  R.~Molina and E.~Oset,
  Phys.\ Rev.\  D {\bf 80}, 114013 (2009)


 
  
\bibitem{BranzGeng}
  T.~Branz, L.~S.~Geng and E.~Oset,
  arXiv:0911.0206 [hep-ph].

\bibitem{raquelnaga}
  R.~Molina, H.~Nagahiro, A.~Hosaka and E.~Oset,
  Phys.\ Rev.\  D {\bf 80}, 014025 (2009)



\bibitem{pdg}
  C.~Amsler {\it et al.}  [Particle Data Group],
  Phys.\ Lett.\  B {\bf 667}, 1 (2008).


\bibitem{mariana}
  M.~Kotulla {\it et al.}  [CBELSA/TAPS Collaboration],
  Phys.\ Rev.\ Lett.\  {\bf 100}, 192302 (2008)
  [arXiv:0802.0989 [nucl-ex]].








\end{thebibliography}
\end{document}